\documentclass[11pt,a4paper]{article}
\usepackage[truedimen,margin=34mm]{geometry}

\usepackage{mathrsfs}
\usepackage{amssymb}
\usepackage{amsmath}
\usepackage{ascmac}
\usepackage{amsthm}
\usepackage[pdftex]{graphicx}
\usepackage[pdftex]{color}

\usepackage{booktabs}
\usepackage{setspace}
\usepackage{natbib}

\usepackage{longtable}

\usepackage{titlesec}
\titleformat*{\section}{\centering\large\bf}
\titleformat*{\subsection}{\centering\it}

\def\ep{{\varepsilon}}

\def\beh{{\widehat \beta}}

\def\pih{{\widehat \pi}}

\def\E{{\rm E}}

\title{{\bf Efficient testing and effect size estimation for set-based genetic association inference via semiparametric multilevel mixture modeling: Application to a genome-wide association study of coronary artery disease}}
\date{}

\begin{document}
\doublespacing

\maketitle

\vspace{-1.5cm}
\begin{center}
{\large Shonosuke Sugasawa$^{1,2*}$, Hisashi Noma$^{2,3}$}
\end{center}

\noindent
$^1$Center for Spatial Information Science, The University of Tokyo, Chiba, Japan\\
$^2$Research Center for Medical and Health Data Science, The Institute of
Statistical Mathematics, Tokyo, Japan\\
$^3$Department of Data Science, The Institute of Statistical Mathematics, Tokyo, Japan

\vspace{3cm}
\noindent
*Correspondence: Shonosuke Sugasawa\\
\hspace{1cm}Center for Spatial Information Science\\
\hspace{1cm}The University of Tokyo\\
\hspace{1cm}5-1-5 Kashiwanoha, Kashiwa, Chiba 277-8568, Japan\\
\hspace{1cm}e-mail: sugasawa@csis.u-tokyo.ac.jp

\newpage
\vspace{0.5cm}
\begin{center}
{\large\bf Abstract}
\end{center}
In genetic association studies, rare variants with extremely small allele frequency play a crucial role in complex traits, and the set-based testing methods that jointly assess the effects of groups of single nucleotide polymorphisms (SNPs) were developed to improve powers for the association tests. However, the powers of these tests are still severely limited due to the extremely small allele frequency, and precise estimations for the effect sizes of individual SNPs are substantially impossible. In this article, we provide an efficient set-based inference framework that addresses the two important issues simultaneously based on a Bayesian semiparametric multilevel mixture model. We propose to use the multilevel hierarchical model that incorporate the variations in set-specific effects and variant-specific effects, and to apply the optimal discovery procedure (ODP) that achieves the largest overall power in multiple significance testing. In addition, we provide Bayesian optimal ``set-based" estimator of the empirical distribution of effect sizes. Efficiency of the proposed methods is demonstrated through application to a genome-wide association study of coronary artery disease (CAD), and through simulation studies. These results suggested there could be a lot of rare variants with large effect sizes for CAD, and the number of significant sets detected by the ODP was much greater than those by existing methods.


\newpage
\section{Introduction}
The advent of high-throughput technologies have realized the whole genome sequencing, and the genome-wide association studies (GWAS) have successfully identified many common single nucleotide polymorphisms (SNPs) associated with complex traits in the last two decades \citep{Manolio2009}. However, a large portion of the heritability of diseases, behaviors, and other phenotypes cannot be accounted by these genetic variations, and it is the well-known the ``missing heritability" problem \citep{Cohen2004}. To explain the missing heritability, recent studies have gradually revealed that ``rare variants" have highly impacts to the complex traits, which are the minor allele frequency (MAF) are smaller than 1-5\% in populations \citep[e.g.][]{Holm2011,Rivas2011}. Besides, detections of the associations of the rare variants are challenging due to their extremely low frequency of rare variants and lacks of the statistical powers for the conventional association tests.

To overcome this underpowered problem, set-based testing methods that jointly assess the associations of a group of SNPs (e.g., a set on a gene, pathway, or network) with a phenotype have been developed in the past decade \citep{Wu2011, Lee2014}.
Various set-based tests for the rare variant analyses based on statistical models were proposed, including C-alpha \citep{Neale2011}, SKAT \citep{Wu2011}, SKAT-O \citep{Lee2014} and HMVD \citep{Cheng2016}.
There have also been several methods summarizing rare variant information within a region into a single genetic score \citep[e.g.][]{LL2008, MB2009, Price2010}.
However, the statistical powers of these set-based tests are still usually insufficient. Also, another problem is the effect size estimation. Due to the extremely low frequency of these variants, the ordinary estimators of effect measures (e.g., odds-ratio) are quite unstable. Existing set-based inference frameworks only provided testing methods, and there have been no effective methods to quantifying the effect sizes of these variants.

In this article, we propose an efficient set-based inference framework that addresses the two important issues simultaneously based on the Bayesian semiparametric multilevel mixture models \citep[e.g.][]{SL1999}. 
We propose to use a multilevel hierarchical model that incorporate the variations in set-specific effects and variant-specific effects, and to apply the optimal discovery procedure (ODP) that achieves the largest overall power in multiple significance testing \citep{Storey2007, NM2012} via the empirical Bayes framework. 
As shown in Section \ref{sec:sim}, the ODP gains the overall powers in the set-based test compared with existing methods such as SKAT-O and HMVD. 
The multiplicity of the set-based ODP can be adequately adjusted with controlling the false discovery rate \citep[FDR;][]{BH1995, Storey2002}.
In addition, the Bayesian formulation enables accurate ``set-based" effect size estimation. It should be noted that the effect size estimation for individual SNPs cannot be suitably implemented, even for the Bayesian shrinkage estimators because unstable ordinary estimators (e.g., maximum likelihood estimator) for the extremely low frequency variants are too shrunken and strongly biased. 
Thus, most of previous rare variant analyses did not discuss the effect size estimates due to these substantial limitations.
However, using the proposed multilevel hierarchical model, we can obtain accurate estimates of effect size distributions of individual units of the set-based analysis alternatively via Bayesian optimal estimation of the empirical distribution function \citep{SL1998}. 
The set-based distribution estimate would provide an additional new relevant information to the set-based inferences.
After presenting these proposed methods, we evaluate their performance by simulation studies and assess their practical usefulness via application to the PennCATH study \citep{Penncath}, a large GWAS for coronary artery disease (CAD).

This article is organized as follows. 
We describe the proposed model model and the ODP in Section \ref{sec:method}.
In Section \ref{sec:sim}, we provide some simulation results, and we apply our approach to the PennCATH study of CAD in Section \ref{sec:app}.
Lastly, we conclude with some discussion in Section \ref{sec:conc}.
Technical details are given in the Appendix.

\section{Methods}\label{sec:method}

\subsection{Hierarchical mixture modeling for effect sizes of multiple rare variants}
Suppose we are interested in group associations of $R$ regions and $M_r$ rare variants are included in the $r$th region ($r=1,\ldots,R$). 
Let $Y_{rk}$ denote a $n_{rk}$-dimensional vector of outcome for the $k$th variant in the $r$th region.
If there are no missing values, $n_{rk}=n$ and $Y_{rk}=Y$. 
Correspondingly, let $X_{rk}$ be a $n_{rk}\times q$ matrix of baseline covariates, and $g_{rk}$ be a $n_{rk}$-dimensional vector of number of minor alleles, so that the elements of $g_{rk}$ are $0, 1$ or $2$.  
We first consider the following generalized linear model:
\begin{equation}\label{model1}
\psi\{\E(Y_{rk})\}=X_{rk}\alpha_{rk}+\beta_{rk}g_{rk}, \ \ \ k=1,\ldots,M_r, \ \ r=1,\ldots,R,
\end{equation}
where $\alpha_{rk}$ is a $q$-dimensional vector of regression coefficient and $\beta_{rk}$ is the (scalar) effect size of the $k$th variant in the $r$th region.
The primary concern in this article is inference on the vector of true effect sizes in each region, $\beta_r=(\beta_{r1},\ldots,\beta_{rM_r})^t$ for $r=1,\ldots,R$.
When $\beta_r=0$, the $r$th region is not associated with the outcome.
On the other hand, the $r$th region is associated with the outcome when at least one element in $\beta_r$ is non-zero.
To express such a structure, we consider the following mixture model for $\beta_r$:
\begin{equation}\label{model2}
\pi(\beta_r)=\pi g_0(\beta_r)+(1-\pi)g_1(\beta_r),
\end{equation}
where $\pi$ is the prior probability of being null, namely, $\pi=P(\beta_{r1}=0,\ldots,\beta_{rM_r}=0)$, and $g_0$ and $g_1$ are null and non-null distributions. 
It is reasonable to force $g_0(\beta_r)$ be the $M_r$-dimensional one-point distribution on $0$.
We propose estimating the non-null distribution $g_1(\cdot)$ in a nonparametric way by incorporating smooth-by-rouging approach \citep{SL1999}, that is, we approximate $g_1(\cdot)$ by the following form:
$$
g_1(\beta_r)=\prod_{k=1}^{M_r}\sum_{\ell=1}^Lp_{\ell}\delta_{a_\ell}(\beta_{rk}),
$$
where $\delta_c(x)$ denotes a one-point distribution on $x=c$, $a_1,\ldots,a_L$ are fixed knots specified by the user and $p_1,\ldots,p_L$ are probabilities of $\beta_{rk}=a_\ell$, satisfying $\sum_{\ell=1}^Lp_{\ell}=1$.
The probabilities $p_{\ell}$'s will be estimated later.

Combination of (\ref{model1}) and (\ref{model2}) gives the posterior distribution of $\beta=(\beta_1^t,\ldots,\beta_R^t)^t$ given by
$$
\pi_{\text{pos}}(\beta;\Theta)=\prod_{r=1}^R\pi(\beta_r;\Theta)\prod_{k=1}^{M_r}\int L(\beta_{rk},\alpha_{rk}|Y_{rk},X_{rk}){\rm d}\alpha_{rk},
$$
where $L(\beta_{rk},\alpha_{rk}|Y_{rk},X_{rk})$ is the likelihood function of (\ref{model1}), and $\Theta=(\pi,p_1,\ldots,p_L)^t$ is a vector of unknown parameters in (\ref{model2}).
Since $L(\beta_{rk},\alpha_{rk}|Y_{rk},X_{rk})$ may have complicated forms, the computation of the integral appeared in the above distribution would be computationally intensive.
Alternatively, we consider an approximated method for computing the posterior distribution of $\beta$.
Specifically, we approximate $\int L(\beta_{rk},\alpha_{rk}|Y_{rk},X_{rk}){\rm d}\alpha_{rk}$ as a function of $\beta_{rk}$ by a normal distribution with mean and variance corresponding to the maximum likelihood estimate $\beh_{rk}$ and its asymptotic variance $s_{rk}^2$ of $\beta_{rk}$ based on the model (\ref{model1}).
The approximation is valid under relatively large $n_{rk}$, which would be often satisfied in practice.  
Therefore, the approximated posterior distribution of $\beta$ is given by
\begin{equation}\label{apos}
\pi_{\text{apos}}(\beta;\Theta)=\prod_{r=1}^R\pi(\beta_r;\Theta)\prod_{k=1}^{M_r}\phi(\beh_{rk}; \beta_{rk},s_{rk}^2).
\end{equation}
This approximated posterior distribution can be derived from the following multilevel models:
\begin{equation}\label{hie}
\begin{split}
\beh_{rk}|\beta_{rk}\sim N(\beta_{rk},s_{rk}^2), \ \ \ k=1,\ldots,M_r, \\
\pi(\beta_r)=\pi g_0(\beta_r)+(1-\pi)g_1(\beta_r),
\end{split}
\end{equation}
independently for $r=1,\ldots,R$. 
In what follows, we call the model (\ref{hie}) semiparametric multilevel mixture model. 
Under the model (\ref{hie}), $\beh_1,\ldots,\beh_R$ are mutually independent and the marginal distribution of $\beh_r$ is given by
\begin{align*}
f(\beh_r; \Theta)
&=\pi\prod_{k=1}^{M_r}\phi(\beh_{rk}; 0, s_{rk}^2)+(1-\pi)\prod_{k=1}^{M_r}\sum_{\ell=1}^Lp_\ell\phi(\beh_{rk}; a_\ell,s_{rk}^2)\\
&\equiv  \pi f_0(\beh_r)+(1-\pi)f_1(\beh_r; P),
\end{align*} 
where $P=(p_1,\ldots,p_L)^t$ is the mixing probabilities in $g_1(\beta_r)$.
The unknown model parameters $\Theta=(\pi,P^t)^t$ can be estimated by maximizing the marginal likelihood function, the product of $f(\beh_r; \Theta)$ for $r=1,\ldots,R$, and the maximization can be efficiently carried out via EM-algorithm \citep{Demp1977}, where its details are provided in the Appendix.

\subsection{Region-specific and variant-specific indices}
Some region-specific indices are useful for screening regions. 
Let $z_r$ be the indicator variable for null/non-null status for the $r$th region, such that $z_r=1$ if the $r$th region is non-null and $z_r$ otherwise. 
The value of $z_r$ is unknown and has the prior probability $P(z_r=1)=1-\pi$. 
Let $D=\{\beh_{rk},s_{rk}, \ k=1,\ldots,M_r, r=1,\ldots,R\}$.
The following posterior probability of being non-null is useful for screening regions:
$$
P(z_k=1|D; \Theta)=\frac{(1-\pi)f_1(\beh_r;P)}{\pi f_0(\beh)+(1-\pi)f_1(\beh_r;P)}.
$$
The effect sizes of each variant $\beta_{rk}$ can be estimated by the posterior mean given by 
$$
E[\beta_{rk}|D;\Theta]=P(z_k=1|D; \Theta)\sum_{\ell=1}^La_\ell P(w_{rk}=\ell|z_k=1,D; \Theta),
$$ 
where 
$$
P(w_{rk}=\ell|z_k=1,D; \Theta)=\frac{ p_\ell\phi(\beh_{rk}; a_\ell,s_{rk}^2)}{\sum_{\ell=1}^Lp_\ell\phi(\beh_{rk}; a_\ell,s_{rk}^2)} \ \equiv \gamma_{rk\ell}(\Theta).
$$
The distribution of effect sizes, $\beta_{r1},\ldots,\beta_{rM_r}$ in each region would provide more valuable and interpretable information than the point estimates. 
In this case, the histogram of posterior means is not necessarily a good estimator of the true distribution \citep{SL1998}, and the optimal estimator of the distribution function is given by
\begin{equation}\label{Dist}
\begin{split}
H_r(t;& \Theta)\equiv
\frac{1}{M_r}\sum_{k=1}^{M_r}P(\beta_{rk}\leq t|D; \Theta)\\
&=I(t\geq 0)P(z_r=0|D;\Theta)+\frac{P(z_r=1|D;\Theta)}{M_r}\sum_{k=1}^{M_r}\sum_{\ell=1}^LI(a_\ell\leq t)\gamma_{rk\ell}(\Theta).
\end{split}
\end{equation}
The posterior indices can be estimated by replacing $\Theta$ with $\widehat{\Theta}$ in (\ref{Dist}).

\subsection{Screening regions based on the optimal discovery procedure}
For detecting and ranking multiple group associations, we follow the optimal discovery procedure \citep{Storey2007, SDL2007} and select significant regions by controlling false discovery rate.
To this end, we use the following statistics for each region induced from the estimated semiparametric multilevel mixture model:
\begin{equation}\label{ODP}
{\rm ODP}_r=\frac{f_1(\beh_r;\widehat{P})}{f_0(\beh_r)}, \ \ \ \ r=1,\ldots,R.
\end{equation}
This is a model-based version of the optimal discovery statistic \citep{Cao2009,NM2012,NM2013}.
For each $r$, we define an index set $I_r$ in which ODP statistics are equal or greater than ${\rm ODP}_r$.
For $I_r$, we can evaluate the model-based false discovery rate \citep[e.g.][]{Mac2006},
$$
{\rm FDR}_r=\frac{1}{|I_r|}\sum_{j\in I_r}P(z_r=0|D;\widehat{\Theta}),
$$
We identify the optimal $r$ whose ${\rm FDR}_r$ is maximum and smaller than pre-specified proportion of false discovery, e.g. $5\%$. 
Due to the flexibility of the semiparametric modeling of the non-null distribution, the proposed method would adequately control the false discovery rate in a wide range of underlying true structures of effect sizes.

\section{Simulation study}\label{sec:sim}
We assessed the performance of the proposed method through simulation studies. 
We considered $n=2000$ individuals, $R=500$ regions.
We randomly generated the number of variants $M_r$ in each region from $1+[\Gamma(5,1)]$, where $\Gamma(a,b)$ denotes the Gamma distribution and $[\cdot]$ is the round function. 
Then, $M=\sum_{r=1}^RM_r$ is the total number of rare variants, which were around 3000 in simulations. 
The minor allele frequency (MAF) of each variant was generated from the uniform distribution on $[0.005,0.01]$.
For generating genotype data, we first generated two $M$-dimensional binary vectors $a_1$ and $a_2$ using \verb+rmvbin+ function in R with a correlation matrix $R=(\rho^{|i-j|})_{i,j=1,\ldots,M}$ with $\rho=0.1$.
We then set the genotype vector $(g_{11},\ldots,g_{1M_1},g_{21},\ldots,g_{RM_R})$ to $a_1+a_2$.  
Also we generated two clinical covariates, $X_1$ and $X_2$ from the standard normal distribution and the Bernoulli distribution with success probability $0.5$.
We considered two cases of response variables and used the following generating model:
\begin{align*}
{\rm (continuous)}& \ \ Y=\gamma_1X_1+\gamma_2X_2+\sum_{r=1}^R\sum_{k=1}^{M_r}\beta_{rk}g_{rk}+\ep, \ \ \ \ep\sim N(0,1)\\
{\rm (binary)}& \ \ Y\sim {\rm Ber}\{e^\eta/(1+e^\eta)\}, \ \ \eta=\gamma_1X_1+\gamma_2X_2+\sum_{r=1}^R\sum_{k=1}^{M_r}\beta_{rk}g_{rk},
\end{align*}
where $\gamma_1=\gamma_2=0.5$.
For the effect sizes of $\beta_r=(\beta_{r1},\ldots,\beta_{rMr})^t$ in each region, we randomly sampled from the following distribution:
\begin{align*}
&\pi(\beta_r)=\pi g_0(\beta_r)+(1-\pi)\prod_{k=1}^{M_r}g(\beta_{rk}), \\
&g(x)=0.9\phi(x; 0.5, (0.2)^2)+0.1\phi(x; -0.3,(0.1)^2),
\end{align*}
where we considered two cases of null probability $\pi$, $\pi=0.5$ and $0.7$.

For the simulated dataset, we applied the proposed ODP method with knots $(a_1,\ldots,a_L)=(-1.00,-0.99,\ldots,0.99,1.00)$.  
As the alternative screening methods, we applied the optimal sequential kernel association test \citep[SKAT-O;][]{Lee2014} and the association test based on hidden Markov model \citep[HMVD;][]{Cheng2016} to each region to calculate $p$-value, both of which are available as R packages.
Then, significant regions are selected using the $q$-value method \citep{ST2003} with controlling the FDR.
Based on 200 simulations, we calculated the average number of detected regions and true positives at FDR$=$5, 10, 15, or 20\% for the two types of responses, which are summarized in Table \ref{tab:sim}. 
Overall, larger numbers of significant regions and true positives are discovered by the proposed method, which indicate its efficiency, compared with the direct use of $q$-values combined with the existing methods.
It is also confirmed that the efficiency gain of the proposed method appeared for smaller values of $\pi$ and larger values of FDR.

\begin{table}[htbp]
\caption{Average numbers of significant biomarkers and true positives for the proposed and alternative methods based on 200 simulations.
\label{tab:sim}}
\begin{center}
\begin{tabular}{ccccccccccc}
\hline
& \multicolumn{4}{c}{\# significant regions}&&\multicolumn{4}{c}{\# true positive} \\
FDR levels & 5\% & 10\% & 15\% & 20\% && 5\% & 10\% & 15\% & 20\%\\
 \hline
Continuous ($\pi=0.7$) \\
ODP (proposed) & 107.8 & 131.2 & 149.9 & 167.4 &  & 101.2 & 116.3 & 124.9 & 131.0 \\
SKAT-O ($q$-value) & 81.8 & 102.9 & 118.8 & 133.9 &  & 78.1 & 93.8 & 103.5 & 110.7 \\
HMVD ($q$-value) & 89.7 & 110.9 & 128.1 & 144.8 &  & 84.6 & 98.9 & 107.9 & 114.5 \\
\hline
Continuous ($\pi=0.5$) \\
ODP (proposed) & 166.2 & 214.4 & 251.6 & 285.0 &  & 155.7 & 188.8 & 208.7 & 221.9 \\
SKAT-O ($q$-value) & 106.8 & 146.0 & 176.9 & 205.6 &  & 102.2 & 134.2 & 155.6 & 172.2 \\
HMVD ($q$-value) & 122.8 & 162.7 & 193.6 & 222.1 &  & 116.5 & 146.7 & 166.0 & 180.8 \\
\hline
Binary ($\pi=0.7$) \\
ODP (proposed) & 27.5 & 50.5 & 72.0 & 93.2 &  & 26.6 & 46.9 & 63.4 & 77.5 \\
SKAT-O ($q$-value) & 16.0 & 30.3 & 44.6 & 59.4 &  & 15.3 & 27.8 & 39.1 & 49.5 \\
HMVD ($q$-value) & 24.9 & 43.4 & 59.4 & 75.6 &  & 23.5 & 38.9 & 50.5 & 60.7 \\
\hline
Binary ($\pi=0.5$) \\
ODP (proposed) & 53.6 & 109.5 & 161.5 & 211.1 &  & 51.6 & 99.6 & 138.0 & 168.5 \\
SKAT-O ($q$-value) & 17.9 & 41.1 & 65.1 & 89.2 &  & 17.3 & 38.6 & 58.7 & 77.4 \\
qHMVD ($q$-value) & 31.8 & 61.7 & 90.9 & 117.9 &  & 30.6 & 56.9 & 80.2 & 99.4 \\
\hline
\end{tabular}
\end{center}
\end{table}

\section{Application}\label{sec:app}
To illustrate the proposed methods in practice, we analyzed a GWAS dataset from the PennCATH study \citep{Penncath}. The PennCATH study is a hospital-based cohort study to evaluate genetic risk factors and biomarkers for CAD, and a nested case-control GWAS study of angiographic CAD was conducted \citep{Penncath}. After the same preprocessing with \cite{Reed2015}, we involved data on $933$ cases and $468$ controls ($1401$ individuals in total) for the GWAS analyses. For the rare variant analyses, we selected the variants whose MAFs were smaller than $5\%$ and at least 3 individuals had the corresponding alleles. We defined the sets of analyses by the groups of SNPs on the same protein coding genes. 
Then, we adopted genetic regions who have at least 10 rare variants, which results in $1094$ regions with total $24902$ SNPs variants in the analyses.

\begin{figure}[!htb]
\centering
\includegraphics[width=14cm,clip]{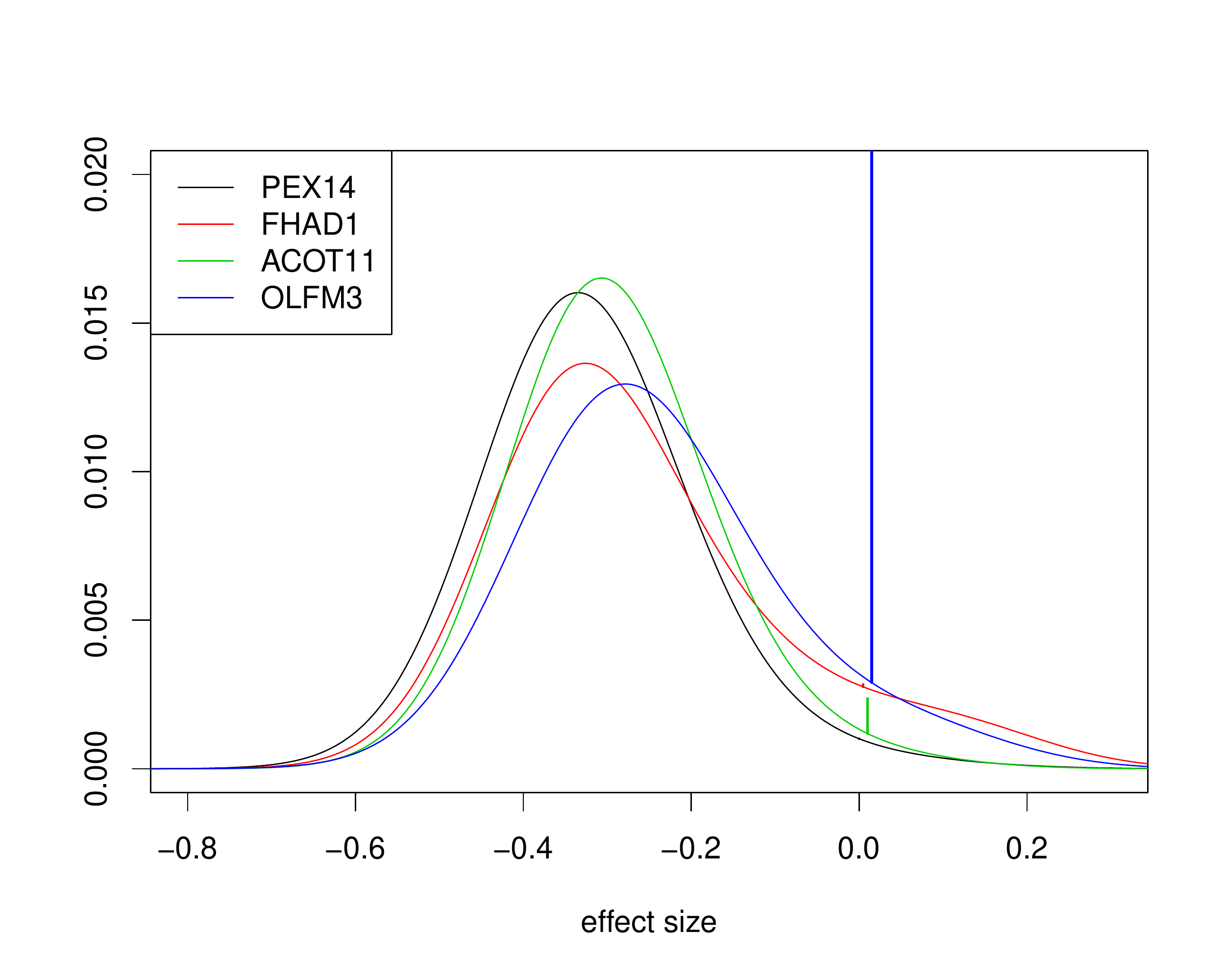}
\caption{The estimated region-wise distribution of effect sizes in selected four selected regions. The bold line on the origin represents the distribution of null effects of rare variants.
\label{fig:Penn-dist}
}
\end{figure}

We fitted the proposed multilevel model by the logistic regression model where the outcome variable was the disease status (case $=1$, control $=0$), and four covariates, age, sex (male $=1$, female $=0$), high- and low-density lipoprotein cholesterol are involved as adjustment variables. Then, we applied the proposed multilevel mixture model with the knots $\{-1.000, -0.995, \ldots ,0.995, 1.000\}$. By the EM algorithm, the estimated null-probability was $\pih=0.861$.
The histogram of z-scores $\beh_{rk}/s_{rk}$ and estimated non-null distribution are shown in Figure \ref{fig:Penn-dist}, which shows that almost all the non-null variants have negative effect sizes. Also, a large fraction (about $70\%$) of the non-null variants was indicated to have log OR smaller than $-0.20$ (for OR scale, about $0.80$).
In addition, $4\%$ of non-null variants were estimated to have log ORs smaller than $-0.50$ (for OR scale, about $0.60$), thus possibly have strong preventive factors of CAD.

For the set-based testing analyses, we first applied SKAT-O and HMVD methods. However, applying the $q$-value method for controlling FDR \citep{ST2003}, no significant regions were detected under FDR control at $5\%$. Even relaxing the threshold to FDR at 20$\%$, SKAT-O detected only 3 significant regions. On the other hand, applying the proposed method, we could detect 74 regions under FDR control at $5\%$. The number of significant sets was much greater than those of the existing methods. The detailed descriptions of the detected regions were provided in the Supplementary Materials.

In addition, using the Bayesian optimal distribution function estimator given by (\ref{Dist}), we estimated the effect size distributions for SNPs of each region.
In Figure~\ref{fig:Penn-dist}, we show the estimated distribution functions in four selected regions which corresponds to ones having the largest upper $5\%$ quantile, the smallest lower $5\%$ quantile, the largest and smallest standard deviations of the estimated distribution functions, respectively.
The estimated frequencies of the null component were different among these regions, and these estimates indicated certain small proportions of SNPs might have large preventive effects ($\log {\rm OR}<-0.5$) for CAD. Also, in the region OLFM3, in spite of the large proportion of null variants, the proposed method could effectively detect such region.
Conventional SNP-specific ML and Bayes estimates cannot provide useful snapshots of these effect size information, so these distribution estimates would provide useful summary for assessing the potential impacts of the detected variants.

\begin{figure}[!htb]
\centering
\includegraphics[width=14cm,clip]{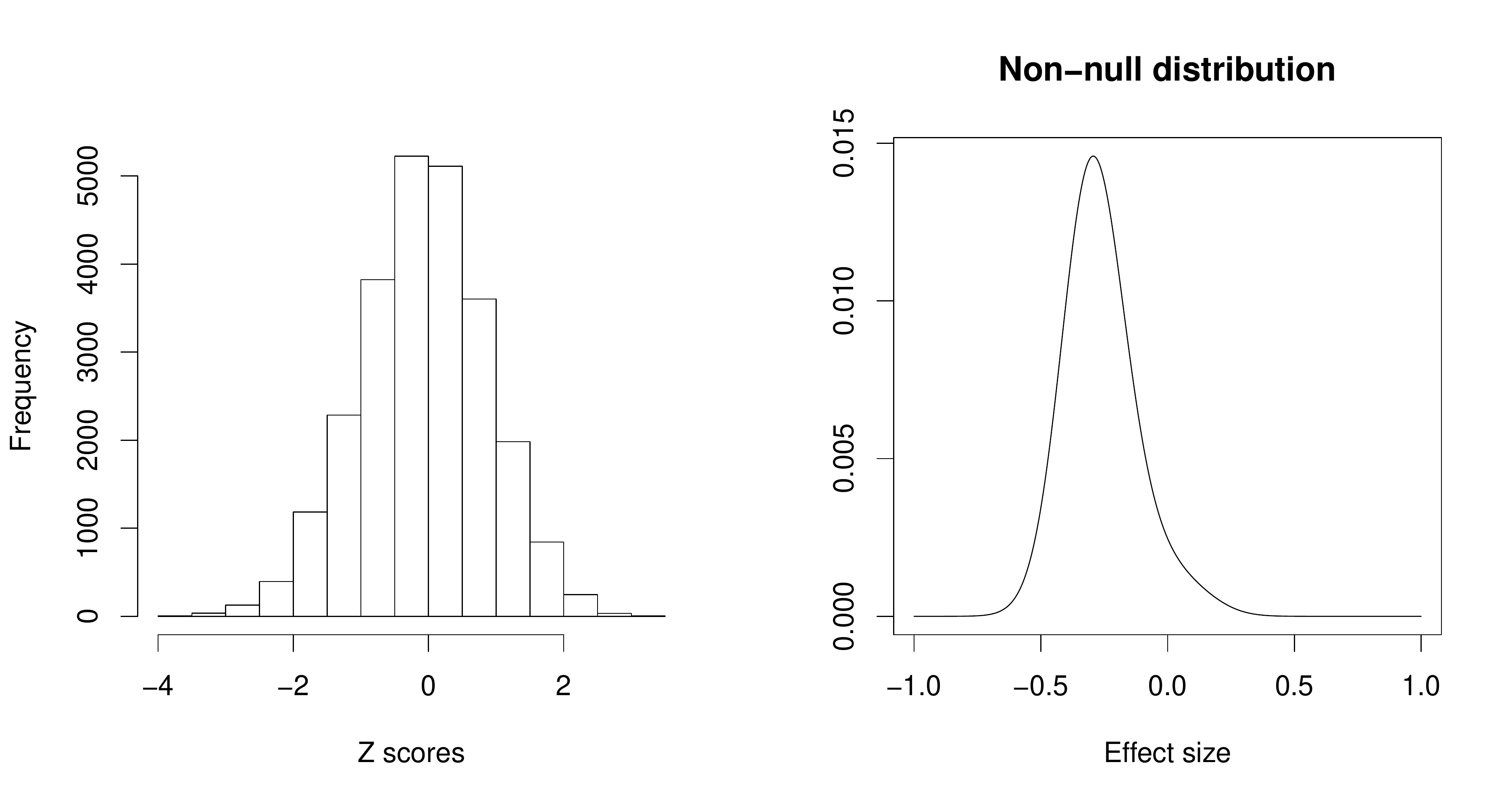}
\caption{Histgram of $z$ scores of all the variants (left) and estimated distribution of the non-null distribution of effect sizes (right).}
\label{fig:Penn-dist2}
\end{figure}

\section{Discussion}\label{sec:conc}
In this article, we developed effective set-based testing and estimating methods for rare variant analyses. As show in the simulation studies of Section \ref{sec:sim}, we could show the overall powers of the set-based tests were improved by the theoretically most powerful testing method, ODP. In addition, using the multilevel hierarchical mixture model, the empirical Bayes inference could borrow the strengths of the among- and within-variants information. These advantages were also reflected to the effect size estimation method, the Bayesian estimator of empirical distribution function of the variant-specific effect sizes. Since one of the main purpose of the rare variant analyses is screening of candidate SNPs for further investigations, the effect size information would be relevant information for the prioritizing steps.

In the applications to PennCATH study, we could detect 74 significant regions under FDR at $5\%$. The number of significant regions were much greater than those of the common rare variant analysis procedures with SKAT-O and HMVD. The discordance of these methods would reflect the relative performances of them in the simulation studies in Section \ref{sec:sim}. Along with the existing evidence for the performances of ODP \citep[e.g.][]{Storey2007, Matsui2018, Otani2018}, the proposed method would have greater overall powers under practical situations.

Besides, the results of the effect size distribution estimation indicated there exists a large number of variants with strong preventive effects of CAD. There have no effective methods that provide effective summary of effect size information in rare variant analyses, thus the proposed new framework would provide relevant information for these studies. Also, although these results indicate the existence of a lot of preventive factors of CAD, the existing methods (i.e., SKAT-O and HMVD) could not detect these factors by the limitations of statistical powers. The proposed ODP also still has the limitation of power, but the effect size estimates provide an alternative effective information to the testing results how these rare variants potentially influences to the heritability of diseases, behaviors, and other phenotypes. The power analyses using this framework would be possible \citep{MN2011BM}. These new information might provide effective hints for the missing heritability problem, and might derive new insights via applying existing rare variant analyses. These practical investigations are subjects for further researches.

\bigskip
\section*{Acknowledgements}

This research was supported by CREST (grant number: JPMJCR1412) from JST, JSPS KAKENHI (grant numbers: JP16H07406, JP18K12757, JP17H01557), and the Practical Research for Innovative Cancer Control (grant number: 17ck0106266) from the Japan Agency for Medical Research.

\vspace{0.5cm}
\section*{Appendix: EM algorithm for computing estimates of the model parameters}

\medskip
\noindent
By introducing latent variables $z_r$ and $w_{rk}$, the multilevel model (\ref{hie}) can be expressed as
\begin{equation*}
\begin{split}
&\beh_{rk}|z_r=0\sim N(0,s_{rk}^2), \ \ \ \ \beh_{rk}|(z_r=1,w_{rk}=\ell)\sim N(a_\ell,s_{rk}^2)\\
&z_r\sim {\rm Ber}(1-\pi), \ \ \ P(w_{rk}=\ell)=p_\ell, \ \ \ell=1,\ldots,L.
\end{split}
\end{equation*}
The complete log-likelihood given the latent variables $z_r$'s and $w_{rk}$'s is given by
\begin{align*}
\log L^c(\Theta)
&=\sum_{r=1}^R (1-z_r)\left\{\sum_{k=1}^{M_r}\log \phi(\beh_{rk}; 0,s_{rk}^2)+\log\pi\right\}\\
& \ \ \ \ 
+\sum_{r=1}^{R}z_r\left[\sum_{k=1}^{M_r}\sum_{\ell=1}^LI(w_{rk}=\ell)\Big\{\log \phi(\beh_{rk}; a_\ell,s_{rk}^2)+\log p_\ell\Big\}+\log(1-\pi)\right].
\end{align*}
By taking the expectation with respect to the latent variables, we have the following objective function in the M-step:
\begin{align*}
Q(\Theta|\Theta^{(t)})
&=\sum_{r=1}^R (1-\xi_r^{(t)})\left\{\sum_{k=1}^{M_r}\log \phi(\beh_{rk}; 0,s_{rk}^2)+\log\pi\right\}\\
& \ \ \ \ 
+\sum_{r=1}^{R}\xi_r^{(t)}\left[\sum_{k=1}^{M_r}\sum_{\ell=1}^L\gamma_{rk\ell}^{(t)}\Big\{\log \phi(\beh_{rk}; a_\ell,s_{rk}^2)+\log p_\ell\Big\}+\log(1-\pi)\right],
\end{align*}
where $\xi_r^{(t)}=P(z_r=1|D;\Theta^{(t)})$ and $\gamma_{rk\ell}^{(t)}=P(w_{rk}=\ell|z_r=1,D;\Theta^{(t)})$. 
The maximization steps of $Q(\Theta|\Theta^{(t)})$ reduces to the updating steps given by 
\begin{align*}
\pi^{(t+1)}=\frac1R\sum_{r=1}^R(1-\xi_r^{(t)}), \ \ \ \ \ \ 
p_{\ell}^{(t+1)}
=\frac{\sum_{r=1}^{R}\sum_{k=1}^{M_r}\xi_r^{(t)}\gamma_{rk\ell}^{(t)}}
{\sum_{r=1}^{R}M_r\xi_r^{(t)}}, \ \ \ \ \ell=1,\ldots,L.
\end{align*} 
Hence, the EM algorithm requires computing $\xi_r^{(t)}$ and $\gamma_{rk\ell}^{(t)}$ in the E-step and update $\pi$ and $p_\ell$'s in the M-step until convergence.

\vspace{0.5cm}
\bibliographystyle{chicago}
\bibliography{ODP-RV}

\newpage

\begin{center}
{\Large \bf Supplementary Material for ``Screening of Multiple Group Associations via Semiparametric Hierarchical Mixture Modeling: Application to Rare Variants Detection"}
\end{center}

\setcounter{equation}{0}
\renewcommand{\theequation}{S\arabic{equation}}
\setcounter{section}{0}
\renewcommand{\thesection}{S\arabic{section}}
\setcounter{table}{0}
\renewcommand{\thetable}{S\arabic{table}}
\setcounter{page}{1}

{\footnotesize
\begin{longtable}{ccccccccccc}
\caption{  {\small  
List of 74 regions detected by the proposed optimal discovery procedure with FDR$=5\%$, and summary statistics of effect sizes of rare variants in each region, including the number of rare variants ($\#$RV), Q-value for the region, and mean, standard deviation (sd) and five (5\%, 25\%, 50\%, 75\%, 95\%) quantiles of the distribution of effect sizes.
The regions are arranged in the ascending order of Q-values. 
}}
\label{tab:sim}
\\
\hline
Gene & $\#$RV & Q-value & mean & sd & $5\%$ & $25\%$ & $50\%$ & $75\%$ & $95\%$\\ 
\hline
\endfirsthead
\hline
Gene & $\#$RV & Q-value & mean & sd & $5\%$ & $25\%$ & $50\%$ & $75\%$ & $95\%$\\ 
\hline
\endhead
 \hline
\endfoot
 \hline
\endlastfoot
RALGAPA2 & 29 & 7.62$\times 10^{-12}$ & -0.292 & 0.142 & -0.495 & -0.385 & -0.305 & -0.215 & -0.035 \\
PKN2 & 16 & 5.40$\times 10^{-9}$ & -0.316 & 0.134 & -0.525 & -0.405 & -0.320 & -0.235 & -0.090 \\
LRRTM4 & 44 & 8.53$\times 10^{-9}$ & -0.266 & 0.158 & -0.500 & -0.370 & -0.280 & -0.175 & 0.025 \\
ABCC4 & 33 & 1.94$\times 10^{-8}$ & -0.279 & 0.149 & -0.500 & -0.380 & -0.290 & -0.195 & -0.010 \\
MCTP1 & 28 & 4.61$\times 10^{-8}$ & -0.280 & 0.147 & -0.495 & -0.380 & -0.290 & -0.195 & -0.015 \\
RSU1 & 31 & 9.23$\times 10^{-7}$ & -0.270 & 0.154 & -0.500 & -0.370 & -0.285 & -0.185 & 0.015 \\
MAN1A1 & 13 & 5.44$\times 10^{-6}$ & -0.301 & 0.133 & -0.500 & -0.390 & -0.310 & -0.220 & -0.070 \\
PTPRK & 56 & 9.04$\times 10^{-6}$ & -0.269 & 0.151 & -0.490 & -0.370 & -0.285 & -0.185 & 0.005 \\
DPP10 & 18 & 1.20$\times 10^{-5}$ & -0.279 & 0.150 & -0.500 & -0.385 & -0.295 & -0.195 & 0.000 \\
C1orf21 & 15 & 1.90$\times 10^{-5}$ & -0.258 & 0.179 & -0.505 & -0.380 & -0.290 & -0.165 & 0.100 \\
MKL2 & 19 & 2.68$\times 10^{-5}$ & -0.271 & 0.149 & -0.495 & -0.370 & -0.285 & -0.185 & 0.000 \\
DIAPH3 & 18 & 3.56$\times 10^{-5}$ & -0.275 & 0.135 & -0.480 & -0.365 & -0.285 & -0.199 & -0.035 \\
GAS7 & 20 & 4.76$\times 10^{-5}$ & -0.278 & 0.151 & -0.505 & -0.380 & -0.290 & -0.190 & -0.005 \\
EPB41L4B & 14 & 6.05$\times 10^{-5}$ & -0.276 & 0.153 & -0.500 & -0.380 & -0.290 & -0.190 & 0.005 \\
SLC24A3 & 61 & 8.08$\times 10^{-5}$ & -0.259 & 0.153 & -0.480 & -0.360 & -0.275 & -0.175 & 0.025 \\
UTRN & 46 & 1.04$\times 10^{-4}$ & -0.268 & 0.152 & -0.495 & -0.370 & -0.280 & -0.185 & 0.010 \\
EPHB2 & 20 & 1.44$\times 10^{-4}$ & -0.285 & 0.148 & -0.510 & -0.380 & -0.295 & -0.200 & -0.025 \\
VAV2 & 16 & 1.89$\times 10^{-4}$ & -0.282 & 0.148 & -0.505 & -0.385 & -0.295 & -0.200 & -0.020 \\
CYB5R4 & 11 & 2.42$\times 10^{-4}$ & -0.292 & 0.130 & -0.490 & -0.380 & -0.300 & -0.215 & -0.065 \\
PRKCH & 46 & 2.96$\times 10^{-4}$ & -0.263 & 0.147 & -0.485 & -0.360 & -0.275 & -0.175 & 0.000 \\
SYN3 & 40 & 3.48$\times 10^{-4}$ & -0.255 & 0.159 & -0.490 & -0.365 & -0.270 & -0.165 & 0.040 \\
CACNB2 & 32 & 4.01$\times 10^{-4}$ & -0.267 & 0.154 & -0.495 & -0.370 & -0.280 & -0.180 & 0.015 \\
NTRK2 & 30 & 4.67$\times 10^{-4}$ & -0.241 & 0.168 & -0.475 & -0.360 & -0.260 & -0.150 & 0.085 \\
CGNL1 & 19 & 5.34$\times 10^{-4}$ & -0.275 & 0.143 & -0.490 & -0.375 & -0.285 & -0.195 & -0.020 \\
CEP85L & 11 & 6.01$\times 10^{-4}$ & -0.285 & 0.140 & -0.495 & -0.380 & -0.295 & -0.200 & -0.040 \\
KLHL1 & 26 & 6.78$\times 10^{-4}$ & -0.265 & 0.143 & -0.475 & -0.360 & -0.275 & -0.185 & -0.005 \\
ZNRF3 & 18 & 7.58$\times 10^{-4}$ & -0.277 & 0.145 & -0.495 & -0.375 & -0.285 & -0.190 & -0.020 \\
CAMKMT & 16 & 8.45$\times 10^{-4}$ & -0.292 & 0.143 & -0.510 & -0.385 & -0.300 & -0.210 & -0.040 \\
GRIK4 & 30 & 9.27$\times 10^{-4}$ & -0.276 & 0.144 & -0.490 & -0.370 & -0.290 & -0.190 & -0.015 \\
ANO3 & 18 & 1.03$\times 10^{-3}$ & -0.272 & 0.140 & -0.480 & -0.365 & -0.280 & -0.190 & -0.030 \\
LINGO2 & 67 & 1.14$\times 10^{-3}$ & -0.245 & 0.159 & -0.475 & -0.355 & -0.265 & -0.160 & 0.060 \\
INTS7 & 14 & 1.28$\times 10^{-3}$ & -0.269 & 0.152 & -0.495 & -0.370 & -0.285 & -0.180 & 0.010 \\
PRKD1 & 18 & 1.43$\times 10^{-3}$ & -0.269 & 0.149 & -0.485 & -0.370 & -0.280 & -0.185 & 0.005 \\
PHACTR1 & 51 & 1.58$\times 10^{-3}$ & -0.259 & 0.143 & -0.470 & -0.355 & -0.270 & -0.180 & 0.005 \\
RBM47 & 14 & 1.78$\times 10^{-3}$ & -0.256 & 0.167 & -0.495 & -0.370 & -0.270 & -0.160 & 0.055 \\
CNTLN & 22 & 1.99$\times 10^{-3}$ & -0.274 & 0.147 & -0.495 & -0.370 & -0.285 & -0.190 & -0.015 \\
ZNF618 & 14 & 2.32$\times 10^{-3}$ & -0.269 & 0.151 & -0.490 & -0.370 & -0.285 & -0.185 & 0.010 \\
ITFG1 & 12 & 2.73$\times 10^{-3}$ & -0.273 & 0.128 & -0.465 & -0.360 & -0.280 & -0.195 & -0.055 \\
PIP4K2A & 11 & 3.30$\times 10^{-3}$ & -0.272 & 0.134 & -0.475 & -0.365 & -0.280 & -0.195 & -0.040 \\
RGL1 & 21 & 3.86$\times 10^{-3}$ & -0.261 & 0.136 & -0.465 & -0.355 & -0.270 & -0.180 & -0.020 \\
CCDC146 & 19 & 4.58$\times 10^{-3}$ & -0.265 & 0.158 & -0.490 & -0.370 & -0.285 & -0.185 & 0.040 \\
DIP2C & 31 & 5.46$\times 10^{-3}$ & -0.236 & 0.162 & -0.475 & -0.345 & -0.250 & -0.145 & 0.065 \\
PID1 & 18 & 6.38$\times 10^{-3}$ & -0.260 & 0.165 & -0.495 & -0.370 & -0.275 & -0.175 & 0.060 \\
NR6A1 & 14 & 7.30$\times 10^{-3}$ & -0.285 & 0.144 & -0.500 & -0.380 & -0.295 & -0.200 & -0.030 \\
UNC5C & 37 & 8.25$\times 10^{-3}$ & -0.253 & 0.154 & -0.480 & -0.355 & -0.265 & -0.165 & 0.035 \\
PTGFRN & 21 & 9.24$\times 10^{-3}$ & -0.262 & 0.138 & -0.465 & -0.355 & -0.275 & -0.180 & -0.015 \\
RAPGEF4 & 24 & 1.02$\times 10^{-2}$ & -0.264 & 0.150 & -0.490 & -0.365 & -0.275 & -0.180 & 0.010 \\
NREP & 12 & 1.13$\times 10^{-2}$ & -0.275 & 0.152 & -0.500 & -0.375 & -0.290 & -0.190 & 0.000 \\
IGF1R & 31 & 1.24$\times 10^{-2}$ & -0.256 & 0.150 & -0.475 & -0.355 & -0.270 & -0.175 & 0.025 \\
MYCBP2 & 12 & 1.35$\times 10^{-2}$ & -0.263 & 0.151 & -0.485 & -0.365 & -0.280 & -0.180 & 0.020 \\
NR3C2 & 18 & 1.48$\times 10^{-2}$ & -0.270 & 0.149 & -0.495 & -0.370 & -0.285 & -0.190 & 0.000 \\
BCAS3 & 30 & 1.62$\times 10^{-2}$ & -0.265 & 0.152 & -0.490 & -0.370 & -0.280 & -0.180 & 0.010 \\
CACNA2D3 & 45 & 1.75$\times 10^{-2}$ & -0.257 & 0.151 & -0.480 & -0.355 & -0.270 & -0.175 & 0.020 \\
TSHZ2 & 16 & 1.89$\times 10^{-2}$ & -0.261 & 0.158 & -0.495 & -0.365 & -0.275 & -0.170 & 0.035 \\
MSR1 & 17 & 2.02$\times 10^{-2}$ & -0.263 & 0.143 & -0.475 & -0.360 & -0.275 & -0.185 & -0.005 \\
ICA1 & 22 & 2.16$\times 10^{-2}$ & -0.254 & 0.155 & -0.480 & -0.360 & -0.270 & -0.170 & 0.040 \\
ADAMTS17 & 43 & 2.29$\times 10^{-2}$ & -0.247 & 0.160 & -0.480 & -0.355 & -0.265 & -0.155 & 0.050 \\
KCNQ5 & 41 & 2.43$\times 10^{-2}$ & -0.253 & 0.149 & -0.470 & -0.350 & -0.265 & -0.170 & 0.020 \\
VAV3 & 16 & 2.55$\times 10^{-2}$ & -0.258 & 0.152 & -0.485 & -0.360 & -0.270 & -0.170 & 0.020 \\
HECTD4 & 12 & 2.69$\times 10^{-2}$ & -0.291 & 0.145 & -0.510 & -0.385 & -0.300 & -0.210 & -0.030 \\
VPS13D & 11 & 2.83$\times 10^{-2}$ & -0.257 & 0.155 & -0.475 & -0.360 & -0.275 & -0.180 & 0.040 \\
LDLRAD3 & 29 & 2.99$\times 10^{-2}$ & -0.254 & 0.153 & -0.480 & -0.355 & -0.270 & -0.170 & 0.025 \\
UVRAG & 23 & 3.15$\times 10^{-2}$ & -0.263 & 0.140 & -0.475 & -0.355 & -0.270 & -0.180 & -0.015 \\
MPPED2 & 10 & 3.31$\times 10^{-2}$ & -0.276 & 0.141 & -0.485 & -0.370 & -0.290 & -0.195 & -0.025 \\
CREBBP & 20 & 3.47$\times 10^{-2}$ & -0.268 & 0.141 & -0.475 & -0.360 & -0.280 & -0.185 & -0.020 \\
NTM & 11 & 3.62$\times 10^{-2}$ & -0.268 & 0.136 & -0.470 & -0.360 & -0.275 & -0.190 & -0.025 \\
MAN1C1 & 12 & 3.78$\times 10^{-2}$ & -0.277 & 0.149 & -0.495 & -0.380 & -0.290 & -0.190 & -0.010 \\
ARHGAP22 & 13 & 3.95$\times 10^{-2}$ & -0.274 & 0.143 & -0.490 & -0.370 & -0.285 & -0.195 & -0.015 \\
SMARCA2 & 18 & 4.12$\times 10^{-2}$ & -0.264 & 0.145 & -0.480 & -0.360 & -0.275 & -0.180 & -0.005 \\
RRP15 & 10 & 4.29$\times 10^{-2}$ & -0.268 & 0.137 & -0.475 & -0.355 & -0.280 & -0.190 & -0.025 \\
MAMDC2 & 11 & 4.45$\times 10^{-2}$ & -0.277 & 0.149 & -0.500 & -0.380 & -0.285 & -0.190 & -0.010 \\
GCOM1 & 17 & 4.63$\times 10^{-2}$ & -0.268 & 0.145 & -0.485 & -0.365 & -0.280 & -0.185 & -0.005 \\
JPH1 & 15 & 4.81$\times 10^{-2}$ & -0.267 & 0.143 & -0.480 & -0.360 & -0.280 & -0.185 & -0.010 \\
LY75-CD302 & 11 & 5.00$\times 10^{-2}$ & -0.263 & 0.144 & -0.475 & -0.360 & -0.275 & -0.180 & 0.000 \\
\end{longtable}
}

\end{document}